\documentclass[aps,prl,groupedaddress,twocolumn]{revtex4}

\usepackage{graphicx}
\usepackage{amsmath}
\usepackage{amsfonts,amssymb}
\newcommand{\ket}[1]{| #1 \rangle}

\newcommand{\rb}[1]{\left( #1 \right)}

\newcommand{\beq}{\begin{eqnarray}}
\newcommand{\eeq}{\end{eqnarray}}

\begin{document}


\title{Entanglement and the Phase Transition in Single Mode Superradiance}


\author{Neill Lambert$^1$, Clive Emary$^2$, and  Tobias Brandes$^1$}
\affiliation{$^1$ Department of Physics,
           UMIST,
           P.O. Box 88,
           Manchester
           M60 1QD,
          England\\
%
$^2$ Instituut--Lorentz,
              Universiteit Leiden,
              P. O. Box 9506 RA Leiden,
              The Netherlands}


\date{\today}

\begin{abstract}
We consider the entanglement properties of the quantum phase
transition in the single-mode superradiance model,
involving the interaction of a boson mode and an
ensemble  of atoms.  For infinite system size, the
atom-field
entanglement of formation diverges logarithmically with the
correlation length exponent. Using a continuous variable representation,
we compare this to the
divergence of the entropy in
conformal field theories, and derive an exact expression for the
scaled concurrence and the cusp-like non-analyticity of the
momentum squeezing.
\end{abstract}


\maketitle
Entanglement has shot to prominence in recent years on the back of
the success of three key areas: quantum computing, quantum
cryptography, and quantum teleportation.  In this quantum
information paradigm, entanglement is a resource which can be
exploited to perform hitherto unimagined physical tasks.

Latterly, a new emphasis has emerged in which entanglement is
related to properties of  interacting many-body systems. This
approach is being pursued most vigourously in connection with
quantum phase transitions (QPTs) \cite{Sachdev}, as it is hoped that entanglement
may shed light upon the dramatic effects occurring in critical
systems which, by their very nature, involve complex collective
quantum mechanical behaviour.  A complete theory of many-body
entanglement is still lacking. Current techniques are reliant upon
bipartite decompositions of the total system, and the criteria
for selecting the most pertinent decomposition are by no means
clear.

Investigations so far have therefore been restricted to
interacting spin-$1/2$ systems on a one-dimensional lattice
\cite{Ostetal02,ON02,VLRK03,Latorre03} or on a simplex \cite{VPM03},
which require
the (more or less artificial) splitting into two spin-subsystems.

In this Letter, we study the entanglement properties of the
one-mode superradiance (Dicke) model \cite{dicke}, where collective and
coherent behaviour of pseudo-spins (atoms) is induced by coupling
(with interaction constant $\lambda$) to a {\em physically distinct}
single-boson subsystem.
We present here exact solutions for the entanglement of formation
between these two subsystems, and for the pairwise entanglement
between atoms at and away from the critical point $\lambda_c$.
Recently the QPT in this model has been related to the emergence
of chaos for $\lambda>\lambda_c$ in a corresponding classical
Hamiltonian \cite{EB03}. 
Our real-space representation of
the modes allows us to analyse the scaling of the atom-field
entanglement {\em at} the critical point, and to compare with results from
conformal field theories for one-dimensional  spin chains
\cite{VLRK03}. Furthermore, we derive explicit expressions for the
concurrence and the related (momentum) squeezing for all coupling
parameters $\lambda$.

A model that has drawn considerable interest in the context
of entanglement near criticality is the $XY$-model. In
ferromagnetic spin $1/2$ chains, the concurrence as a function of
system size been used \cite{Ostetal02} to demonstrate scaling of
entanglement near the transition point. Osterloh {\it et al.}
\cite{Ostetal02} have shown that the derivatives of the
concurrence between neighbour and next-nearest neighbour spins
exhibits a universal scaling behaviour in the region of the
critical point in this model. Furthermore, the study of such
systems has led Osbourne and Nielsen
\cite{ON02}
to the notion of a `critically entangled' system where the correlation length $\xi$
of the system is divergent and entanglement exists over all length
scales.
Vidal {\em et al.}
\cite{VLRK03}
have used an alternative approach and studied the entanglement
between blocks of $L$ contiguous spins and the rest of the chain
and have found a striking relation to the entropy $S_L\approx
(c+\bar{c})/6\log L + const$ in $1+1$ conformal field theories
with central charges $c$ and $\bar{c}$.

We start by describing our model, which is
the single-mode Dicke Hamiltonian describing
the interaction of $N$ two-level atoms of splitting $\omega_0$
with a single bosonic mode of frequency $\omega$
\beq
\cal{H} &=&
   \omega_0 \sum_{i=1}^N s_z^{(i)}
    + \omega  a^\dagger a
    + \sum_{i=1}^N
        \frac{\lambda}{\sqrt{N}} \rb{a^\dagger + a}
        \rb{s^{(i)}_+ + s^{(i)}_-}
  \nonumber \\
  &=&
  \omega_0 J_z + \omega a^\dagger a
  + \frac{\lambda}{\sqrt{2j}} \rb{a^\dagger + a}\rb{J_+ + J_-},
\label{DHam1}
\eeq
where the second form follows from the introduction of
collective spin operators of length $j=N/2$.
In the thermodynamic limit, $N, j \rightarrow \infty$,
the system
undergoes a QPT at a critical coupling of
$\lambda = \lambda_c=\sqrt{\omega \omega_0}/2$, at which
point the system changes from
a largely unexcited normal phase to a super-radiant one in which
both the field and atomic collection  acquire macroscopic occupations.

Similar to the large-spin problem analysed in this context
\cite{VPM03}, the Dicke Hamiltonian can be regarded as a
zero-dimensional field theory with mean-field type behaviour, where
the $S_N$ permutation symmetry of the atoms and the absence of an
intrinsic length scale makes the model exactly solvable.
Despite this simplicity,
our model exhibits  many non-trivial
properties; in particular, exact solutions for the
non-analyticities of the entanglement of formation and the
concurrence can be related to the scaling exponent, the
finite-size behavior, and the underlying semi-classical
integrable/chaos cross-over which has been shown to occur around
the phase transition \cite{EB03}.

\begin{figure}[t]
\centerline{
  \includegraphics[clip=false,width=\columnwidth]
{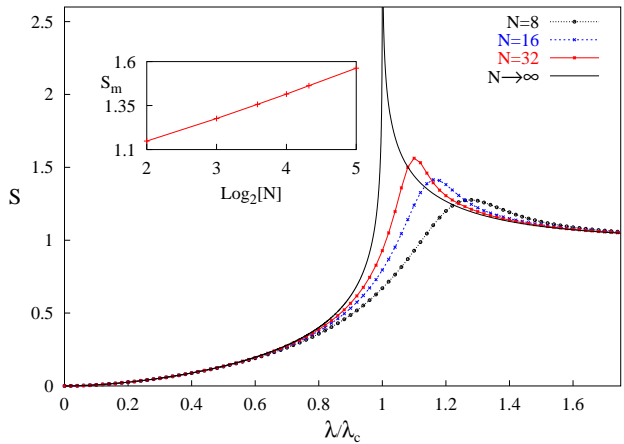}
} \caption{\label{entropy} Entanglement of formation $S_{\infty}$
between atoms and field for both $N \rightarrow \infty$ and finite
$N$. Inset: Scaling of the value of the entanglement maximum as a
function of $\log_2N$. The Hamiltonian is on scaled resonance
$\omega =\omega_0 = 1$.}
\end{figure}

The starting point for our analysis in the thermodynamic limit
is the Holstein-Primakoff
representation \cite{HP49} of the angular momentum operators
$J_z = \rb{b^\dagger b - j}$,
$J_+ = b^\dagger \sqrt{2j - b^\dagger b}$,
$J_- = J_+^\dag$.
Here,  $b$ and $b^\dagger$ are bosonic operators that convert
$\cal{H}$ into a two-mode bosonic problem.  This allows us to obtain
effective
Hamiltonians
that are exact in the thermodynamic limit,
by neglecting terms from expansions of the Holstein-Primakoff
square-roots\cite{EB03}.
In the normal phase, $\lambda < \lambda_c$, we expand the
square-roots directly and obtain the effective Hamiltonian
\beq
  {\cal{H}}^{(1)} = \omega_0 b^\dagger b + \omega a^\dagger a
  + \lambda \rb{a^\dagger + a}\rb{b^\dagger + b} - j \omega_0.
  \label{lcDH1}
\eeq In the super-radiant phase, we first displace both boson
modes by quantities proportional to $\sqrt{j}$ before we
approximate the square-roots.  This leads to a second effective
Hamiltonian, the form of which is also bilinear and similar
to Eq. (\ref{lcDH1}).

We now consider the normal phase ground state in some detail; the
super-radiant phase results following with slight modification.
The eigenstates of ${\cal H}^{(1)}$ are two-mode squeezed states.
Via the introduction of a position-momentum representation for the
two oscillators; $x \equiv \frac{1}{\sqrt{2 \omega}}\rb{a^\dagger
+ a}$, $y \equiv\frac{1}{\sqrt{2 \omega_0}}\rb{b^\dagger + b}$,
with the momenta defined canonically, we may write the
ground-state wavefunction as \beq
  \Psi(x,y)=
  \rb{\frac{\varepsilon_+ \varepsilon_-}{\pi^2}}^{\frac{1}{4}}
  e^{
    -\frac{\varepsilon_-}{2}\rb{c x - s y}^2
    -\frac{\varepsilon_+}{2}\rb{s x + c y}^2
    }
  \label{wfn}
\eeq
where
$\varepsilon_{\pm}^{^2} = \frac{1}{2} \rb{\omega^2 + \omega_0^2 \pm
\sqrt{(\omega_0^2 - \omega^2)^2 +16\lambda^2\omega \omega_0}}$ are the
excitation energies of the system,
$s\equiv \sin \gamma$, $c \equiv \cos \gamma$, and the angle
$2\gamma = \arctan[4 \lambda \sqrt{\omega\omega_0}/
  {(\omega_0^2 - \omega^2)}]$
characterises the squeezing axis. This wavefunction forms the
basis of the current analysis.

{\it Entanglement of formation}.-- As a measure of the
entanglement between the atoms and the field, we calculate the
von-Neumann entropy $ S \equiv -\mbox{\rm tr}\hat{\rho} \log_2
\hat{\rho}$ of the reduced density matrix (RDM) $\hat{\rho}$ of
the field-mode. 
In the normal
phase, $\hat{\rho}$ is simply determined by the ground state wave
function, Eq. (\ref{wfn}), whereas in the super-radiant phase {\em
two} degenerate ground states exist that have wave functions
$\Psi_{\pm}$ similar to Eq. (\ref{wfn}), but displaced from the
origin by amounts proportional to $\pm \sqrt{j}$. This degeneracy
arises from the breaking of the parity symmetry $\Pi=
\exp\left\{i\pi \left[a^\dagger a + J_z+j \right]\right\}$ for
$\lambda>\lambda_c$. Because $\Psi_+$ and $\Psi_-$ are orthogonal,
the convex nature of the von-Neumann entropy \cite{Wehrl78}
implies that in the SR phase
$S(\hat{\rho}_{\rm cat}) = S(\hat{\rho}_{\pm}) + 1$, where
$\hat{\rho}_{\pm}$ is the RDM of either of the two
(macroscopically separated for large $N$) solutions, and
$\hat{\rho}_{\rm cat}$ is the RDM of the superposition `cat' state
of the two.  The cat state restores the broken parity, and thus
the latter expression will be used for comparison with the
numerical results for finite $N$.

Having clarified this additional distinction between the two
phases, we now explicitly calculate the normal phase RDM in the
$x$-representation,
\begin{eqnarray}\label{rhoL}
  \rho_L(x,x')=c_L\int_{-\infty}^{\infty}dy f_L(y)\Psi^*\rb{x,y}\Psi\rb{x',y}.
\end{eqnarray}
Here, $c_L$ is a normalisation constant, and the introduction of
the cut-off function $f_L(y)\equiv e^{-y^2/L^2}$ will allow us to
discuss the effect of a partial trace over the atomic ($y$) modes
(see below). A straightforward calculation shows that $\rho_L$ is
identical to the density matrix of a single harmonic oscillator
with frequency $\Omega_L$ in a canonical ensemble at temperature
$T\equiv1/\beta$, where
\begin{eqnarray}
\label{cosh}
  \cosh \beta \Omega_L = 1 + 2\frac{\varepsilon_-\varepsilon_+ + 4(\varepsilon_-c^2+\varepsilon_+s^2)/L^2}
{(\varepsilon_--\varepsilon_+)^2c^2s^2}.
\end{eqnarray}
The entropy $S_L$ obtained from $\rho_L$ is thus given by the expression
\beq
  S_L(\zeta) = \left[\zeta
\coth \zeta - \ln (2\sinh \zeta)\right]/\ln 2,\quad \zeta\equiv
\beta\Omega_L/2.
  \label{sT}
\eeq This strikingly simple result allows some interesting
observations.
First of all, the entropy $S_\infty$ (cut-off $L=\infty$) undergoes a divergence at the
critical point as we approach $\lambda_c$ from either side. In the
region near $\lambda_c$, the excitation energy $\varepsilon_-$
vanishes as $\varepsilon_- \propto |\lambda-\lambda_c|^{2 \nu}$,
with the exponent $\nu =1/4$ describing the divergence of the
characteristic length $\xi \equiv \varepsilon_-^{-1/2}$. Using
$S_\infty(\zeta)=[1-\ln (2\zeta)+\zeta^2/6]/\ln 2+O(\zeta^4)$ and $\zeta=
\sqrt{\varepsilon_{\infty}/2}[1+O(\varepsilon_{\infty})]$ with
$\varepsilon_{\infty}\equiv 2\varepsilon_-/(\varepsilon_+s^2c^2)$,
we find that $S_{\infty}$ diverges logarithmically as $S_{\infty}
\propto -(1/2)\log_2(2\varepsilon_{\infty})$ and hence (omitting
constants),
\begin{eqnarray}
S_{\infty} \propto
- \nu \log_2 |\lambda-\lambda_c| = \log_2 \xi, \quad \nu = 1/4.
\end{eqnarray}
Thus, the entanglement between the atoms and field
diverges with the same critical exponent as the
characteristic length - a clear demonstration of critical
entanglement.

As we approach $\lambda_c$ the  parameter
$\zeta=\hbar \Omega_{\infty}/k_B T$ of the fictitious thermal
oscillator approaches zero, indicating that a {\em classical}
limit of the field RDM is being approached,
interpreted either as the temperature $T$ going to infinity, or
the frequency $\Omega_{\infty}$ approaching zero. In terms of the
original parameters of the system, the dependence of the entropy
is through the ratio of energies
$\varepsilon_{\infty}\propto\varepsilon_-/\varepsilon_+$\cite{Sre93}.
Although the entanglement calculated here is a
genuine quantum property of the combined atom-field system, the
above results highlight that in the limit of $N\to \infty$ atoms,
the exact mapping of the system to two coupled oscillators leads
to emergent pseudo-classical behaviour. This is consistent with
the observation that the Dicke Hamiltonian in fact is strongly connected to a
classical (cusp) singularity in the catastophe theory sense
\cite{EB03b}.

We next compare the analytical result from
Eq.(\ref{cosh},\ref{sT}) for the entropy $S_{\infty}$
(corresponding to completely tracing out the atomic mode) with
the corresponding finite $N$ results obtained from numerical
diagonalisation. Fig. 1 shows these results, and illustrates the
finite size scaling. Defining $\lambda^\mathrm{M}$ as the position
of the entropy maximum, and $S_\mathrm{M}$ as the value of the
maximum entropy, we observe $\lambda^\mathrm{M} - \lambda_c
\propto N^{-0.75 \pm 0.1}$, and $S_\mathrm{M} \propto (0.14 \pm
0.01) \log_2N$.

The accuracy of the exponents are limited by the available
numerical data.  
{The divergence of the entropy is logarithmic due to the symmetric nature of the spin system. The entropy here saturates at a maximum value of $\log_2 (N+1)$, in contrast with general spin sytems which saturate at $\log_2 (2^N)$ due to their larger Hilbert spaces}. 
This distinction is
expected to be important in determining the complexity of
classically simulating a quantum phase transition
\cite{Vidal203,Orus03,Latorre03}.  An explicit plot of the entropy
scaling is shown as an inset in Fig. 1, while the scaling of the
position of the maximum point is shown as an inset in Fig. 2.

We next consider the system {\em at} the critical point
but keep the `tracing parameter' $L$ finite. This
corresponds to a situation where the trace over the (atomic)
$y$-coordinate is performed over only a finite gaussian effective
region of size $L$ for the atomic wave function. With
$\varepsilon_-=0$, the relevant dimensionless energy scale is now
$\varepsilon_L\equiv 2/(L^2\varepsilon_+ c^2)$, and the
entanglement entropy diverges as (again omitting constants)
\begin{eqnarray}\label{SLinf}
S_L\propto -(1/2)\log_2(2\varepsilon_L)= \log_2L,\quad L\to \infty.
\end{eqnarray}
This result can now be compared with a recent calculation by Vidal {\em et al.}
\cite{VLRK03} of the critical
entanglement of formation of blocks of $L$ spins in one-dimensional
interacting $XY$ and $XXZ$ spin-chain models.
There, the prefactor for the $\log L$ dependence of $S_L$ at criticality
is given by the central charges of the underlying
conformal field theory in $1+1$ dimensions.
Note, however, that a direct comparison would require the
tracing out of $L$ atoms from the $N$-atom Hamiltonian (see below) with
$N \to \infty$, $L$ fixed, but the general principle is the same.
In this context,
the Dicke model corresponds to a
zero-dimensional field theory and, for $N\to \infty$,
is in fact closely related to
Srednicki's  simple two-oscillator model
in his introductory discussion of entropy and area \cite{Sre93}.

{\it Pairwise entanglement and concurrence}.--
\begin{figure}[t]
\centerline{
  \includegraphics[clip=false,width=\columnwidth]{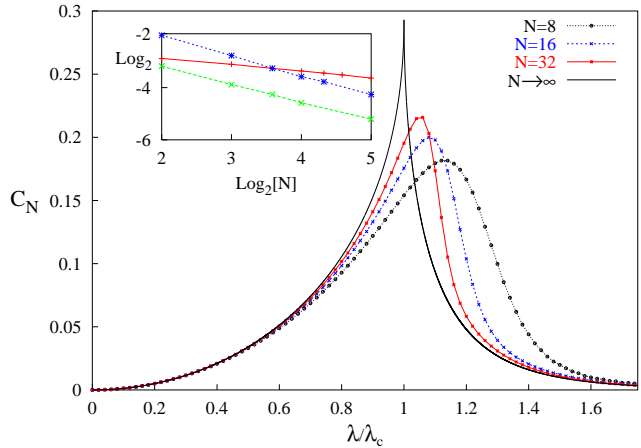} 
} \caption{\label{pair} Scaled pairwise concurrence  $C_N=NC$
between two spins for both $N \rightarrow \infty$ and  finite $N$.
Inset:  Scaling of the value ($+$) and position ($\times$) of the
concurrence maximum, and the position of the entropy maximum ($*$)
as functions of $N$. The Hamiltonian is on scaled resonance
$\omega =\omega_0 = 1$.}
\end{figure}
To observe the behaviour of the entanglement between atoms within
the ensemble, we proceed by considering the `pairwise'
entanglement of formation for mixed states, as
parameterized by the concurrence \cite{Wootters98}. 
The absence of an intrinsic
length scale in our model simplifies our calculations, enabling us
to employ the prescription set out for symmetric Dicke states in
\cite{WM02}. The matrix elements of the reduced density matrix
$\rho_{12}$ for any two atoms is determined by the expectation
values of the collective operators, $\langle J_z \rangle$,
$\langle J_z^2 \rangle$, and $\langle J_+^2 \rangle$. 
We then
define the {\em scaled} concurrence
as $C_N\equiv N C$, with
$C\equiv{\rm max} \{0,\lambda_1-\lambda_2-\lambda_3-\lambda_4\}$,
where the $\lambda_i$ are the square roots of the eigenvalues (in
descending order) of $\rho_{12}(\sigma_{1y}\otimes \sigma_{2y})
\rho_{12}^* (\sigma_{1y}\otimes \sigma_{2y})$. 
Wang and Sanders  have shown that $C_N$ can be expressed in 
terms of the  Kitagawa-Ueda
spin squeezing \cite{Kitagawa} for symmetric multi-spin 
states\cite{Wang&Sanders}.

We show numerical results for the scaled concurrence $C_N$ in Fig. 2,
together with the analytic thermodynamic limit result described below.
For all $\lambda$ and $N$, $C_N$
is less than that of the pure $W$--state $\ket{j,\pm (j-1)}$,
which has $C_N=2$, the maximum pairwise concurrence
of any Dicke state \cite{WM02}.
For small coupling $\lambda$, we recognise an $N$-independent
behaviour of $C_N$ which may be obtained from
perturbation theory in $\lambda$ as
\begin{eqnarray}
C_N(\lambda\to 0) \sim 2\alpha^2/(1+\alpha^2),\quad \alpha\equiv
\lambda/(\omega+\omega_0).
\end{eqnarray}
As with the entropy, we can perform a finite scaling analysis of
the numerical data.  Again, two power law expressions are found
for $\lambda^\mathrm{M}$ and $C^\mathrm{M}_{N}$; $
\lambda^\mathrm{M} - \lambda_c \propto N^{-0.68 \pm 0.1}$ and
$C^\mathrm{M}_N(\lambda_c) - C_N \propto N^{-0.25 \pm 0.01}$.
Plots of this behaviour are shown as an inset in Fig. 2.

In the thermodynamic limit $N\to \infty$, the scaled concurrence
can be expressed as
\begin{eqnarray}\label{Cs}
  C_{\infty} = (1+\mu)\left[\langle (d^{\dagger})^2 \rangle -
  \langle d^{\dagger}d\rangle \right]
+\frac{1}{2}(1-\mu),
\end{eqnarray}
setting $\mu=1$ and $d^{\dagger}=b^{\dagger}$
in the normal phase ($\lambda<\lambda_c$),  and
$\mu=(\lambda_c/\lambda)^2$
and $d^{\dagger}=b^{\dagger}+\sqrt{N(1-\mu)/2}$ in the SR phase
($\lambda>\lambda_c$).
Recalling $b^{\dagger} = \sqrt{\omega_0/2}(y-ip_y/\omega_0)$,
we can further transform
Eq.(\ref{Cs}) to establish a
relation between the scaled concurrence, the {\em momentum
squeezing} $(\Delta p_y) ^2 \equiv \langle p_y^2 \rangle - \langle
p_y \rangle^2$.  We have
\begin{eqnarray}
C_{\infty} &=& (1+\mu)\left[\frac{1}{2} - (\Delta p_y)^2/\omega_0\right] + \frac{1}{2}(1-\mu),
\end{eqnarray}
where 
again, setting $\mu=(\lambda_c/\lambda)^2$ gives the super-radiant
phase equivalent.
The concurrence can be
explicitely related to the parameters in the reduced $y$ (atom)
oscillator density matrix as
$
C_{\infty}= 1-({\mu \Omega}/{\omega_0})\coth (\beta \Omega/2)
$
with $\cosh \beta \Omega = 1+2\varepsilon_-\varepsilon_+/D$,
$D\equiv [cs(\varepsilon_--\varepsilon_+)]^2$, and $2\Omega/\sinh
\beta\Omega = D/(\varepsilon_-c^2+\varepsilon_+s^2)$. Due to
symmetry, these are the same parameters as for the reduced field
($x$) density matrix $\rho_{\infty}$, Eq. (\ref{rhoL}), with
$s=\sin \gamma$ and $c=\cos \gamma$ interchanged. After simple
algebra one obtains $ C_{\infty} =
1-\mu(\varepsilon_-s^2+\varepsilon_+c^2)/\omega_0$.
{Due to space restrictions, we only give analytical  results
at resonance ($\omega=\omega_0$)},
  \begin{eqnarray}\label{cusp1}
C_{\infty}^{x\le 1}&=&1-\frac{1}{2}\left[\sqrt{1+ x}+\sqrt{1- x}\right],\quad x\equiv \lambda/\lambda_c\\
\label{cusp2}
C_{\infty}^{x\ge 1}&=&1-\frac{1}{\sqrt{2}x^2}\left[
\left(\sin^2 \gamma\right)\sqrt{1+x^4
- \sqrt{ \left(1-x^4\right)^2 + 4}}\right.\nonumber\\
&+&\left.\left(\cos^2 \gamma\right)\sqrt{1+x^4
+ \sqrt{ \left(1-x^4\right)^2 + 4}} \right],
  \end{eqnarray}
where 
$2\gamma=\arctan[2/(x^2-1)]$ in the SR phase.
{These explicit expressions reveal the  square-root non-analyticity
of the scaled concurrence near the critical point $\lambda_c$.}
The concurrence assumes its maximum $C_{\infty} = 1-\sqrt{2}/2\approx 0.293$
{\em at}  the critical point $\lambda=\lambda_{c}$.
We note that  Eq.~(\ref{cusp1}) is consistent
with the maximum of the (unscaled) concurrence approaching the critical point
in a related, dissipative version of the Dicke model in the 
normal phase\cite{SM02}.
Our findings are also in agreement with the
behaviour of the concurrence in the collective spin-model,
$H=-(2\lambda/N)(S_x^2+\gamma S_y^2)-2S_z+(\lambda/2)(1+\gamma)$ \cite{VPM03}
and differ from 1D spin chains, where the maximum of the $C$ does not
coincide with its non-analyticity at the critical point.
We also note here that the squeezing  obtains its minimal
value at $\lambda_c$, which is again in agreement with the above spin model.

In conclusion, we have obtained exact results for the entropy
and the concurrence
in a model that allows us to quantify entanglement across a quantum
phase transition.
The clear physical distinction between the sub-systems
(pseudo-spin or two-level system and  bosonic mode)  enables us
to see distinctly
the logarithmic divergence of the entropy in the thermodynamic
limit as a function of the coupling constant. We mention that
quantum phase transitions have also been discussed very recently
in the context of  entanglement generation
(e.g., for atoms in optical lattices \cite{Doretal03}),
and quantum computation schemes \cite{Orus03}.
A further though still mainly unexplored aspect remains the
fundamental role of the phase transition for the
connection between entanglement and the
underlying integrable to quantum chaotic
transition.
\begin{acknowledgments}
This work was supported by projects EPSRC GR44690/01, DFG
Br1528/4-1, the WE Heraeus foundation, 
the Dutch Science Foundation NWO/FOM, 
and the UK Quantum Circuits Network.
N. L.
acknowledges
discussions with R. Or\'{u}s,
and
the participants at Session LXXIX of the Les Houches Summer
School, 2003.
\end{acknowledgments}


\end{document}